\documentclass[conference]{IEEEtran}
\IEEEoverridecommandlockouts
\usepackage{cite}
\usepackage{amsmath,amssymb,amsfonts}
\usepackage{algorithmic}
\usepackage{graphicx}
\usepackage{textcomp}
\usepackage{xcolor}
\def\BibTeX{{\rm B\kern-.05em{\sc i\kern-.025em b}\kern-.08em
    T\kern-.1667em\lower.7ex\hbox{E}\kern-.125emX}}

\usepackage[frozencache,cachedir=.]{minted}
\newcommand{\cd}[1]{\mintinline[fontsize=\small ,escapeinside=@@,mathescape=true]{java}{#1}}
\usepackage{tikz}
\usetikzlibrary{arrows,shapes,positioning}

\newcommand{\XHIDE}[1]{}
\newcommand{\Xcensor}[1]{#1}

\begin{document}

\title{Having Fun in Learning Formal Specifications
   \thanks{\Xcensor{Funded by the EU Erasmus{+} grant 2017-1-NL01-KA203-035259.}}
}


\author{
\IEEEauthorblockN{
   \Xcensor{I.S.W.B. Prasetya}
   , \Xcensor{Craig Q.H.D. Leek}
   , \Xcensor{Orestis Melkonian}
   , \Xcensor{Joris ten Tusscher} 
   , \Xcensor{Jan van Bergen} 
   , \Xcensor{J.M. Everink} \\  
   , \Xcensor{Thomas van der Klis}   
   , \Xcensor{Rick Meijerink} 
   , \Xcensor{Roan Oosenbrug} 
   , \Xcensor{Jelle J. Oostveen}  
   , \Xcensor{Tijmen van den Pol} 
   , \Xcensor{Wink M. van Zon} 
   }
\IEEEauthorblockA{ \ \\ \Xcensor{\textit{Dept. of Information \& Computing Sciences, Utrecht University}}}

}

\maketitle

\begin{abstract}

There are many benefits in providing formal specifications for our software. However, teaching
students to do this is not always easy as courses on formal methods are often experienced
as dry by students. This paper presents a game called FormalZ that teachers can use to
introduce some variation in their class. Students can have some fun in playing the game
and, while doing so, also learn the basics of writing formal specifications. Unlike  
existing software engineering themed education games such as Pex and Code Defenders, FormalZ
takes the deep gamification approach where playing gets a more central role in order
to generate more engagement. This short paper presents our work in progress: the first
implementation of FormalZ along with the result of a preliminary users' evaluation. 
This implementation is functionally complete and tested, but the polishing of its user 
interface is still future work.

\end{abstract}


\noindent
\fbox{
\parbox{85mm}{\scriptsize
This is a preprint of: 
Prasetya, Leek, Melkonian, Tusscher, van Bergen, Everink, van der Klis, Meijerink, Oosenbrug, Oostveen, van den Pol, and van Zon,
{\em Having Fun in Learning Formal Specifications},
in the 41st International Conference on Software Engineering (ICSE) 2019 --- Software Engineering Education and Training (SEET) Track,
IEEE, 2019.}
}

\section{Introduction}

In the world of fast churning software industry, we might wonder whether applying formal
methods is a viable option, since formal proofs are arguably hard to produce.
Even if we can get programmers with the needed 
mathematical skill to produce them, the process is too slow to keep up with the pace of modern agile development.
On the bright side, we do not need to exercise the full scale of formal methods to reap its
benefit. Much can already be gained just by writing formal specifications. 
This does not require sophisticated mathematical skills, and nowadays not even a separate
specification language and a separate tool chain anymore. Many modern 
programming languages support $\lambda$-expressions, which allows, for instance,
predicate logic formulas to be expressed natively in the programming languages
themselves.

Having formal specifications enables verification through automated testing 
or bounded model checking\footnote{While one 
can employ these techniques in the absence of a formal specification, only the 
correctness with respect to general properties,
such as absence of crash or abnormal CPU usage, can be verified.}. 
While this benefit might be clear in the eyes of a computer scientist, 
convincing practitioners to write formal specifications is still not easy.
The myth that any form of exercising formal methods requires sophisticated
math remains, and this is perhaps also a message that programmers somehow
picked up from their education, where lectures in formal methods
tend to be terse and dry. Students receive more points from being able
to construct formal proofs. Demonstrating the ability to write formal specifications 
receives less points, hence creating the perception that it is less important (whereas
we just argued that it is a more usable skill).

This paper presents a browser-based education game called FormalZ that teachers can use
to break the typical monotony in a class on formal method by allowing
students to have some fun playing the game, while also
learning the basic of writing formal specifications. Unlike existing software engineering
themed education games like Pex~\cite{tillmann2011pex} and Code Defender~\cite{ICSE_SEET2017_CodeDefenders},
FormalZ takes a deeper gamification approach~\cite{boyce2014deep}, where 'playing' 
is given a more central role. After all, what makes games so engaging
is not merely the awarded scores and badges, but primarily the experience
of playing them. The ultimate research question that intrigues us is whether
such an approach will actually make a difference towards the game's ultimate learning goal.
This is still on-going work. Currently, we have a fully functional implementation of 
FormalZ, but polishing the user interface is still future work.  
In this short paper we will present its game concepts and the result
of a preliminary users' evaluation.


\section{FormalZ Game Concepts} \label{sec.concept}

In FormalZ, teachers provide exercises in the form of program headers and informal descriptions
of the intended program behaviour. For each exercise, the student is asked to translate
the informal description into formal pre- and post-conditions. An example of such an exercise
is shown below:

\begin{quote}\em
"Given a non-empty array $a$, the program \textup{\cd{int getMax(int[] a)}} returns the greatest element in the array."
\end{quote}    

The teacher accompanies the exercise with a solution in the form of a formal specification,
written in a Domain Specific Language (DSL) that closely resembles predicate logic formulas.
The DSL is embedded in Java, so the teacher can simply use Java SDK to test the solution
(if it reflects what he/she has in mind), before deploying the exercise to the class.
Section~\ref{subsec.javadsl} will provide more details on this DSL.
    
\begin{figure}
\begin{center}
\includegraphics[scale=0.18]{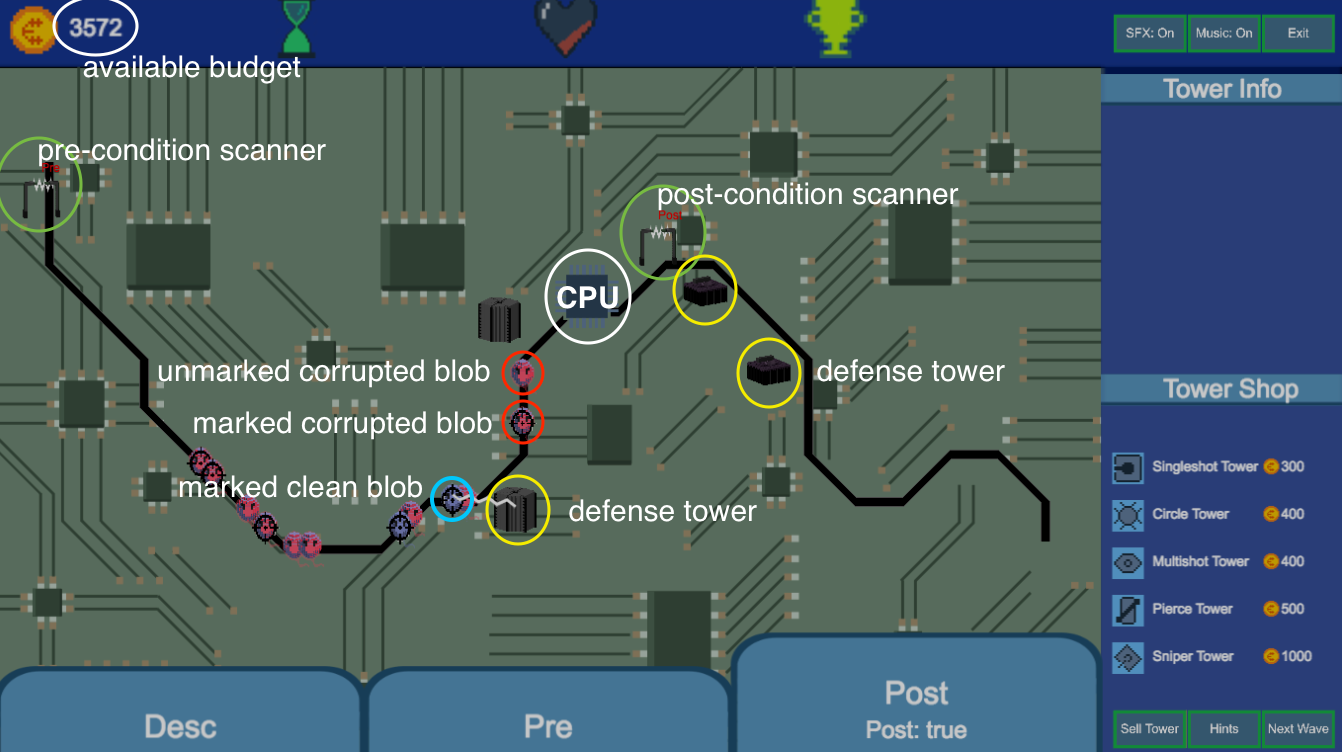} 
\end{center}
\vspace{-3mm}
\caption{\em A screenshot of FormalZ. The CPU that has to be defended is in the middle.
The small red and blue blobs represent data coming to or leaving the CPU. 
The green circles are the pre- and post-condition scanners,
used to mark blobs --- we can see that some red blobs are left unmarked (not good),
and some blue blobs are marked (also not good). Defence towers (yellow circles) must be placed
to shoot down marked blobs; one tower can be seen as 'zapping'
at a wrongly marked blue blob. More towers can be placed to get higher score, subject to
the available budget (top left).
}
\label{fig.gameshot1}
\end{figure}

A student solves the exercise by offering a pre- and post-condition which are equivalent
to the teacher's solution. FormalZ frames this in a game of defending a CPU, which is
a re-interpretation of the popular tower defence genre of games. The CPU symbolizes the
program that is being specified. The story line is that hackers manage to find a way to
influence data packages that flow into and out from the CPU. 
Data packages are represented by blobs, see Figure~\ref{fig.gameshot1}: red blobs
are data that the hackers manage to corrupt to incorrect values, and blue blobs are 
those that are still 'clean'. No incoming red blob should reach the CPU, nor leave the CPU 
to reach the environment.
To defend against this attack, the CPU's circuit board provides two scanners,
one on the input side of the CPU, and one on its output side. Both
can be programmed to identify and mark certain blobs. Other hardware called 'defence towers' can be
added to the circuit board to discard marked blobs. 

The scanners symbolize the pre- and post-conditions that the student should construct.
An incorrectly programmed scanner would leave some red blobs unmarked,
and may also wrongly mark blue blobs. The latter is also bad, since they represent good
inputs or outputs, which the towers would subsequently wrongly discard. 

\subsection{Constructionism}

\begin{figure}
\begin{center}
\includegraphics[scale=0.18]{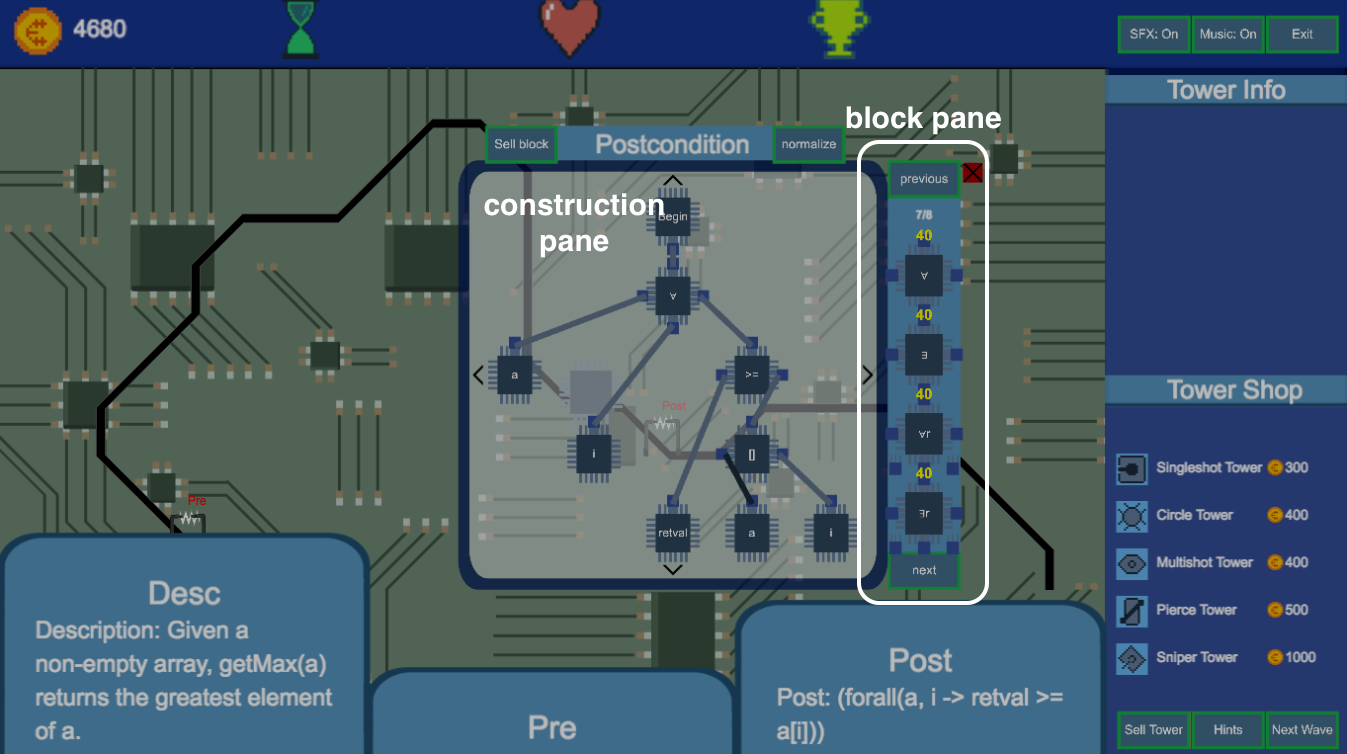} 
\end{center}
\vspace{-3mm}
\caption{\em The construction pane of FormalZ. The user can construct a pre- or post-condition by dragging blocks,
essentially the AST of the formula that the user has in mind.}
\label{fig.gameshot2}
\end{figure}

To construct the pre- and post-conditions the student 
gets a special construction panel where blocks can be placed
to construct a formula; see Figure~\ref{fig.gameshot2}.  
Each block represents either a variable, a constant or
an operator. Wires are used to connect the blocks to construct the tree representation
of the formula; its so-called Abstract Syntax Tree (AST). Simply typing the formula
is deliberately disallowed; 
let us motivate this from the perspective of the Constructionism theory of learning~\cite{papert1991constructionism}.

The theory believes that humans learn by constructing knowledge, rather 
than by simply transferring this knowledge from a teacher into the head of a watching learner. 
Familiar physical objects play a key role in this process,
because the learner already has knowledge on how they work~\cite{KafaiConstructionism05}. When new knowledge is 
framed in terms of interactions with these familiar objects, it helps the learner
to construct the new knowledge in his mind.
The theory was originally proposed by Papert and Harel~\cite{papert1991constructionism}.
Papert used LOGO as an example, which he used to teach programming to children.
A learner can easily relate the 'turtle' in LOGO with his/her own physical 
body which can turn and move forward and draws upon this analogy to learn
LOGO programming concepts.

In FormalZ, the blocks (in the formula
building) are visually depicted as electronic hardware components, which are concepts
that do exist in the physical world (as opposed to e.g. a "variable" which does not really have tangible physical existence) and can be assumed to be recognizable to most computer
science students. Likewise, the wires that connect the blocks relate well to our physical
experience, where electronic components always need to be connected by wires.
By requiring the user to first locate the right block, drag it to the construction
pane and then explicitly connect it to other blocks with wires, we enforce more 
self-conscious interactions by the user, hence creating a more gradual and deliberate
process of knowledge construction, as opposed to just letting the user type in formulas.

\subsection{Deep gamification}

Introducing gamification just by adding game components like points,
badges and leaderboards (what in \cite{boyce2014deep} is called shallow gamification),
would miss some key aspects such as play and fun, which are important to
make a game engaging~\cite{prensky2003ChFun}. For example, awarding
a badge to a student for completing a certain task acknowledges a certain achievement, but
it does not mean that the task is fun to do. Making the task fun 
would evoke more engagement, which we believe will improve learning as well (an opinion also shared
by Prensky in his classic work "{\em{Digital Game-Based Learning}}"~\cite{prensky2003ChFun}). 

Although constructing formulas using blocks is to some degree fun, it is still a quite formal task 
as it has to abide by a whole set of rules (e.g. we cannot connect an \fbox{\footnotesize\sf \&\&} block to 
an integer block, nor connect it to more than one parent) and is therefore not really
'playful'. 
An important characteristic of a play, as Caillois~\cite{caillois2001man} puts it, is that it
is not obligatory; if it were, it will lose its joyous quality. Although playing
FormalZ is not obligatory (at least, we do not envision it to be so), having more
rules does evoke some sense of being obliged to do things in a certain way.
To create play, FormalZ therefore frames the challenge of constructing
pre- and post-conditions as a defence game where the goal 
is to prevent corrupted blobs from reaching their destination, while letting through
as many clean blobs as possible. To do this the player also has to strategically
place defence towers on the circuit board to shoot and remove marked blobs. There are different kinds of towers,
each with unique features. The selection and placement of the towers are subject to
almost no restriction. The student is free to experiment to figure out which strategies lead
to better scores, though ultimately best scores are only attainable with the help of
correct pre- and post-conditions.

\subsection{Java DSL} \label{subsec.javadsl}

To specify pre- and post-conditions in a model solution, the teacher must write them in a DSL.
The main design criterion was easy integration with the Java ecosystem
to allow teachers to test out solutions using the Java SDK 
and enable future extensions where
students are also asked to implement the specified program.
To this end we chose to embed the DSL in Java, rather than follow a more extrinsic approach (e.g. custom Java pre-processor).
%
To specify pre-conditions or post-conditions, we use the \cd{pre} and
\cd{post} functions, supplied with a \cd{boolean} expression as the argument.
These expressions must be Java expressions, constructed using known
boolean operators such as negation (\cd{!}), conjunction (\cd{&&}) and disjunction (\cd{||}).
To aid readability, we allow multiple pre/post statements, which are conjunctively interpreted
(e.g. \cd{pre(f1);pre(f2)} $\equiv$ \cd{pre(f1&&f2)}).

The DSL also allows implication (\cd{imp(f1,f2)} $\equiv$ f1 $\Rightarrow$ f2)
and polymorphic equality (\cd{==}) between a fixed universe of types, namely
integers (\cd{int}, \cd{short}, \cd{long}),
reals (\cd{float}, \cd{double}),
and multi-dimensional arrays of the previous types (e.g. \cd{float[][]}).
Any equality check on expressions of a different type is equivalent to \cd{false}, except when Java semantics allow coercions
between different numeric representations.

The usual representation-agnostic numeric operations are supported: \cd{+},\cd{-},\cd{*},\cd{/},\cd{@\%@}
and comparisons (\cd{<,<=,>,>=}).

To allow formal specifications of programs working on arrays, we additionally support the following array operations:
\begin{itemize}
\item Retrieving an array's length, as in \cd{a.length > 0}.
\item Indexing an array, as in \cd {a[5] == 0}.
\item Universally/existentially quantifying over the elements of an array, as in \cd{forall(a, i -> a[i]==0)}.
\end{itemize}
%
Notice that we use $\lambda$-expression \cd{args->body}, which was added to Java
since Java-8, to model universal/existential quantification.
Below is a possible specification of the exercise shown in Section \ref{sec.concept}:

\begin{minted}
[ frame=lines,
  xleftmargin=5mm,
  xrightmargin=5mm,
  fontsize=\scriptsize 
]{java}
public static void getMax_spec(int[] a) {
  // pre-conditions
  pre(a != null);
  pre(a.length > 0);
  
  // call to actual implementation
  int retval = getMax(a);
  
  // post-conditions
  post(exists(a, i -> a[i] == retval));
  post(forall(a, i -> a[i] <= retval));
}
\end{minted}

\subsection{Checking specifications} \label{subsec.spec}

\XHIDE{
To enable future extension of our system to other programming languages, we
define a simple first-order logic augmented with array operations and null checks. This 
intermediate representation (IR) acts as a common target language, to be targeted
by several programming language frontends.

For brevity's sake, we omit a formal definition of the logic's syntax and semantics.
It suffices to say that this intermediate representation is essentially the same with
the one described in the previous subsection, without the syntactic overhead Java imposes on us.
}

\XHIDE{
With this common interface in place, we can implement several backends that compare
the student's solution against the teacher's model answer.
}

Given a teacher specification $M$ and a student specification $S$, 
FormalZ's backend will try to determine whether the two specifications
are \emph{semantically} equivalent: $ M \equiv S$.
%
%
%
While this is ultimately what the student is aiming for, the backend also provides helpful
feedback in the case of an incorrect student solution. Specifically, when $M \not\equiv S$
it also gives the following information: 
\begin{itemize}
\item If the student solution is too strong: $S \Rightarrow M$
\item If the student solution is too weak: $M \Rightarrow S$
\item Which of these combinations are satisfiable:
$M \land S,\ M \land \neg S,\ \neg M \land S,\ \neg M \land \neg S$. This information
controls the generation of the blobs and their
marking. E.g. if $\neg M \land S$ is satisfiable the game will generate
at least one unmarked red blob, and if  $M \land \neg S$ is satisfiable
then at least one marked blue blob is generated.
\end{itemize}

\XHIDE{
\begin{figure}
\centering
\begin{tikzpicture}[
  base/.style={rounded corners,draw=black,minimum width=1.5cm,text centered},
  align=center,
  scale=0.3
]
\node (top) at (3,3) {};
\node (bottom) at (3,-3) {};
\node[base,fill=yellow!20, font=\scriptsize] (java)  at (-6,0) { Java DSL};
\node[base,fill=gray!20, font=\scriptsize] (logic)   at (0,0) {Logic IR};
\node[base,fill=green!20, font=\scriptsize] (z3)     at (8,1) {\scriptsize Z3};
\node[base,fill=green!20, font=\scriptsize] (random) at (8,-1) {Random testing};

\draw[->] (java) -- (logic);
\draw[->] (logic) -- (z3.west);
\draw[->] (logic) -- (random.west);
\draw[dash pattern=on5pt off3pt] (top) -- (bottom);

\node[left=of top,font=\scriptsize] {Frontend};
\node[right=of top,font=\scriptsize] {Backend};

\end{tikzpicture}
\vspace{-3mm}
\caption{\em Overall architecture of the specification checker}
\label{fig:logicir}
\end{figure}
}
\noindent
FormalZ implements two backends.

\subsubsection*{Z3}

The principal Z3 backend converts the abstract syntax tree (AST) of Java expressions
to the AST used by the Z3 theorem prover~\cite{z3}. We can then freely invoke the Z3 solver to \emph{try} to prove that the queried logical formula is actually true.

\subsubsection*{Random testing}

While most teacher examples are expected to lie in Z3's decidability range, in 
general they are undecidable. So, we also provide a second backend that employs random testing to approximate the aforementioned truth-values.

The random testing backend first looks at $M$ and $S$ to identify clauses that are present in both, possibly in different but equivalent forms (e.g. the teacher wrote $a \Rightarrow b$ whereas the student wrote $\neg a \lor b$). It then removes these clauses, since they cannot possibly influence the validity of $M\equiv S$, and would result in simpler formulas on which an randomized equivalence test can be performed more quickly.

After the elimination step, the random testing backend starts a repeating process. In every iteration of the process, it first generates random values for all variables present in either $M$ or $S$, 
including non-primitive variables like (multidimensional) arrays. It then substitutes these variables in both $M$ and $S$ for the generated random values and evaluates the two formulas. During the evaluation step, universal and existential quantifiers are only partially evaluated, meaning that only a limited number of iterations of the quantifier are actually evaluated, for the sake of running time. After the evaluation step, $M$ and $S$ will reduce to a single boolean literal,
from which the backend concludes whether $M\equiv S$ is satisfied, or if that is not the case, which other cases (see above) do apply. This approach is repeated multiple times to increase the 
confidence,
and the combined results are returned by the backend.

\section{Preliminary Evaluation} \label{sec.evaluation}

\begin{figure*}
	\begin{center}
    \begin{tabular}{p{45mm} p{45mm} p{45mm}}    
        \hfil \includegraphics[scale=0.34]{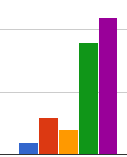} &
        \hfil \includegraphics[scale=0.34]{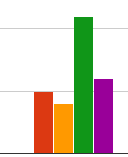} &
        \hfil \includegraphics[scale=0.34]{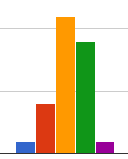} \\
        {\em\footnotesize (1) I understood what the goal of the game was.} &
        {\em\footnotesize (2) The way Tower Defence and Block Building were connected was clear.} &
        {\em\footnotesize (3) Figuring out what I could do was enjoyable.} \\
        \hfil \includegraphics[scale=0.34]{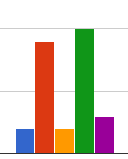} & 
        \hfil \includegraphics[scale=0.34]{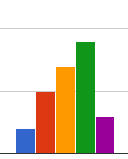} &
        \hfil \includegraphics[scale=0.34]{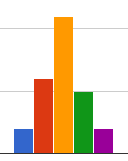} \\
        {\em\footnotesize (4) The way the resources 
          worked was clear, and they were fun to manage.} &
        {\em\footnotesize (5) The provided gameplay feedback was clear and understandable.} &
        {\em\footnotesize (6) I liked the overall look of the game.} \\
        \hfil \includegraphics[scale=0.34]{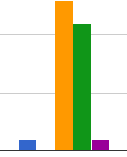} &
        \hfil \includegraphics[scale=0.34]{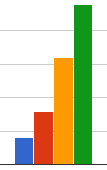} &
        \hfil \includegraphics[scale=0.34]{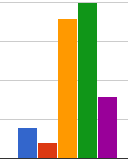} \\
        {\em\footnotesize  (7) I had fun while playing the game.} &
        {\em\footnotesize  (8) Thinking of ways to combine towers and create set-ups was fun.} &
        {\em\footnotesize  (9) Figuring out what blocks to make use of was interesting and enjoyable.}
    \end{tabular}    
	\end{center}
	\caption{\em The evaluation results. Each question takes the form of an assertion, to which
a student can disagree (blue --left most bar), somewhat disagree, neither agree nor disagree,
somewhat agree, or agree (purple --right most bar).
    } \label{fig.results}
\end{figure*}

To evaluate how our concept worked with students we ran a playtest with students that were taking the bachelor course \Xcensor{Software Testing and Verification at Utrecht University}. This group was chosen as this is the typical target audience for the game. 
They played a previous version of the game and feedback received in this session has already been incorporated in the game as it is now. After they played the game for approximately 30 minutes, including a short tutorial they could take to get introduced to the basics of the game, they were asked to complete a questionnaire. This resulted in 26 responses. Key results
are shown in Figure~\ref{fig.results}.
\\
We can see that the ultimate goal of the game was understood by most students (graph 1), as well as a majority being able to see the link between the separate key elements (defence towers and block building, question 2) of the game. However when we went deeper into the different aspects of the game, it was clear that students struggled slightly more with understanding how everything worked. An example of this is question 4 on resource (e.g. money and towers) management where the result shows a clear split: 13 people agreed to an extend, while 11 disagreed. We are not sure what exactly causes this split; it might be related to prior gaming experience of the students. 
%
%
FormalZ's way of giving feedback does resonate with students in general, as they were positive 
on average about it (graph 5). Improvement is still called for, 
as quite a few people also did not think the feedback was clear.
Students were less positive about the overall look of the game (graph 6). This is something that is actively worked on as a priority. 

The students are positive about having room to 
figure things out themselves (that is, to be allowed to 'play', which is a point deep gamification
tries to put more emphasis on), as is shown in  graph 8
for the Tower Defence aspect specifically and graph 9 for the Block Building. 
Overall, graph 7 shows that students are neutral or positive about the enjoyability of the game on average.


\section{Conclusion and Future Work} \label{sec.concl}

We presented a fully functional implementation of the game FormalZ, which teachers can use
as an alternative medium to help them teach students how to write formal 
specifications. FormalZ adopts a combination of the constructionist and the deep-gamification
approaches. Our preliminary evaluation indicated that most our subjects perceived the approach
positively. Despite the greater emphasis on the 'play' element, most subjects at least understood what the game's
goal was. We cannot however confirm yet, if such an approach would indeed improve the students'
learning. This requires further experiments, which for now are left as future work.
Additionally, the game needs some cosmetic improvement to make it look more pleasing and presentable; 
this is also future work.


\section*{Acknowledgement}

We thank \Xcensor{Sergey Sosnovsky} for his useful feedback throughout the development of FormalZ.

\bibliographystyle{IEEEtran}
\bibliography{references}

\end{document}